\documentclass[shortnames,nofooter,noheadings]{jss}
\usepackage{float}
\usepackage[utf8]{inputenc}
\usepackage{graphicx,subfig}
\usepackage{placeins}
\usepackage{fancyhdr}
\restylefloat{figure}

\author{Joseph Stachelek\\   South Florida Water Management District
\And
  Christopher J. Madden}
\title{Application of Inverse Path Distance Weighting for high density spatial mapping of coastal water quality patterns}

\Plainauthor{Joseph Stachelek, Christopher J. Madden} 
\Plaintitle{Application of Inverse Path Distance Weighting for high density spatial mapping of coastal water quality patterns} 
\Shorttitle{IPDW for coastal mapping} 

\Abstract{
\noindent One of the primary goals of coastal water quality monitoring is to characterize spatial variation. Generally, this monitoring takes place at a limited number of fixed sampling points. The alternative sampling methodology explored in this paper involves high density sampling from an onboard flow-through water analysis system (Dataflow). Dataflow has the potential to provide better spatial resolution of water quality features because it generates many closely spaced (< 10 m) measurements. Regardless of the measurement technique, parameter values at unsampled locations must be interpolated from nearby measurement points in order to generate a comprehensive picture of spatial variations. Standard Euclidean interpolations in coastal settings tend to yield inaccurate results because they extend through barriers in the landscape such as peninsulas, islands, and submerged banks. We recently developed a method for non-Euclidean interpolation by inverse path distance weighting (IPDW) in order to account for these barriers. The algorithms were implemented as part of an \proglang{R} package and made available from R repositories.  The combination of IPDW with Dataflow provided more accurate estimates of salinity patterning relative to Euclidean inverse distance weighting (IDW). IPDW was notably more accurate than IDW in the presence of intense spatial gradients.
}
\Plainkeywords{non-Euclidean distance, spatial interpolation, flow-through sampling, salinity, Florida Bay, estuary} 

 \Year{2015}
 \Acceptdate{2015-02-07}

\Address{
  Joseph Stachelek\\
  Everglades Division\\
  South Florida Water Management District\\
  3301 Gun Club Rd, West Palm Beach, USA\\
  E-mail: \email{jstachel@sfwmd.gov}\\
}

\begin{document}


\section[Introduction]{Introduction}
\noindent One of the primary goals of water quality monitoring is to characterize spatial variation. However, the resolution of spatial features can be limited both by the widely-spaced fixed-point design of monitoring programs and by spatially complex landscape features. The alternative sampling methodology explored in this paper involves high density sampling from an onboard flow-through water analysis system (Dataflow). Dataflow \citep{madden1992instrument} has the potential to provide better spatial resolution of water quality features because it generates many closely spaced (<10 m) measurements.

  Regardless of the measurement technique, parameter values at unsampled locations must be interpolated from nearby measurement points in order to generate a comprehensive picture of spatial variations. Interpolation approaches vary in complexity from simple inverse distance techniques to complex kriging algorithms \citep{zimmerman1999experimental}. In many settings, the analyst can simply choose an interpolation technique based on the properties of the sampling network and the spatial dependence of the variable of interest \citep{isaaks1989applied}. In coastal settings, however, interpolations can be confounded by the presence of landscape features such as peninsulas, islands, and submerged banks.  For example, two monitoring stations geographically close to one another may be hydrologically separated by a peninsula. 

  As a result, standard interpolations that extend through barriers and do not account for them are likely to provide inaccurate results. There are several existing methods for dealing with the presence of barriers while using standard Euclidean interpolations. Most notably, these include "interpolation with barriers" whereby a static barrier is introduced in order to isolate disparate portions of the study area \citep{krivoruchko2004geostatistical,soderqvist2010seasonal}. The interpolation with barriers approach is not ideal because points are excluded from the interpolation based on line-of-sight. As a result, the inherent connectivity present in aquatic environments (water flow, diffusion) is not taken into account. 

  We developed an alternative method for interpolation using inverse path distance weighting (IPDW) that honors barriers in the landscape while more accurately accounting for aquatic connectivity because it can "round corners." IPDW belongs to a class of techniques that use "in-water" path distances (non-Euclidean) rather than "as the crow flies" (Euclidean) distances as input to interpolation routines \citep{little1997kriging}. Unfortunately, these techniques have received limited consideration in part because until recently they had not been implemented within existing software tools. In this study, we utilize a software tool that uses path distances as input to inverse distance weighting (IDW). IDW is a deterministic interpolation method that has some disadvantages relative to geostatistical methods such as kriging because there is no model fitting and thus no assessment of prediction error. Path distances have been used as input to kriging with some success \citep{krivoruchko2004geostatistical,lopez2009geostatistical} but software tools are still under development. 

  This research is the first study that utilizes IPDW with a spatially dense water quality dataset generated by Dataflow sampling. We explore the benefits and limitations of this technique using a case study set in Florida Bay, USA. Florida Bay represents an ideal candidate for IPDW because it is a complex array of embayments and fragmented basins (Figure~\ref{fig:one}). In addition, we were able to reliably test interpolation accuracy because of the high measurement density afforded by Dataflow. We specifically compare IPDW with its Euclidean counterpart IDW.
\section[Methods]{Methods}

\noindent \subsection[Field Data Collection]{Field Data Collection}
\noindent Field measurements were collected during shipboard cruises using a Dataflow onboard flow-through collection system. While underway, the Dataflow receives a continuous stream of water from an onboard pump. This water is routed past a series of water quality probes that measure the temperature, dissolved oxygen, specific conductivity, chlorophyll a, and chromophoric dissolved organic matter (cdom) of pumped water at six second intervals. Salinity is calculated from conductivity using the Practical Salinity Scale (PSS-78) and is defined as a unitless ratio \citep{iociapso}. Each measurement is geo-referenced with an integrated global positioning system (GPS) unit. The focus of the present study is on salinity and only salinity measurements are presented hereafter. This choice was made in part because of the ecological importance of salinity in Florida Bay and the fact that these measurements are generally subject to less measurement error than other parameters.

\begin{figure}[h]
\begin{center}
\includegraphics{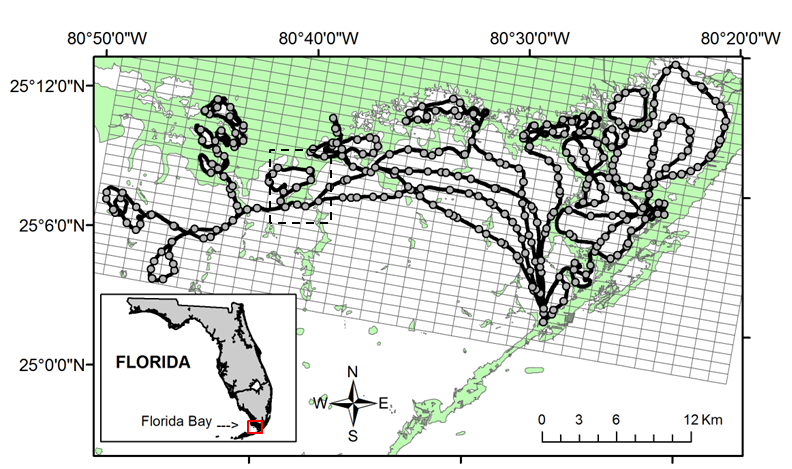}
\end{center}
\vspace{-20pt}
\caption{Map of northern Florida Bay showing the approximate track of Dataflow surveys (solid line), the grid used to subset full Dataflow surveys (grid),  a representative subset of a full survey used for interpolation (filled circles), and location of Figure~\ref{fig:two} (dashed box). The location of Florida Bay on the Florida peninsula is also shown (inset map).}
\label{fig:one}
\end{figure}

  Each survey generated approximately 6,000 data points across an approximately 600 km2 portion of northern Florida Bay, eastern Florida Bay, and southern Biscayne Bay (Figure~\ref{fig:one}). Cruises took place four times per year from 2006-2012 usually twice in the dry season (November – May) and twice in the wet season (June - October). Cruises last several hours and are thus quasi-synoptic. However, measurements were minimally affected by tides because the submerged banks in Florida Bay attenuate most of the tidal signal. The timescale of complete surveys was only a fraction of the most significant tidal signal (14-d long-period lunar component) in the eastern Bay \citep{wang1994wind}. The route of each survey was nearly identical and was intended to capture the influence of freshwater discharge from the Everglades on Florida Bay.
  
\FloatBarrier

\newpage
\subsection[Statistical Analysis]{Statistical Analysis}
\noindent A variant of inverse distance weighting (IDW) called inverse path distance weighting (IPDW) was used in order to account for barrier effects during spatial interpolation \citep{suominen2010surface}. IDW is a deterministic interpolation procedure that estimates values at prediction points (V) using the following equation:

\begin{equation}
\label{eq:one}
V = \frac{\sum\limits_{i=1}^n v_i \frac{1}{d_i^p}}{\sum\limits_{i=1}^n \frac{1}{d_i^p}}
\end{equation}

where d is the distance between prediction and measurement points, vi is the measured parameter value, and p is a power parameter \citep{isaaks1989applied}. The advantage of IPDW is that it uses non-Euclidean "path distances" for d. These path distances are calculated using an algorithm that accounts for the cost of travel from one cell to the next.

\vspace{20pt}
\begin{figure}[h]
\begin{center}
\includegraphics{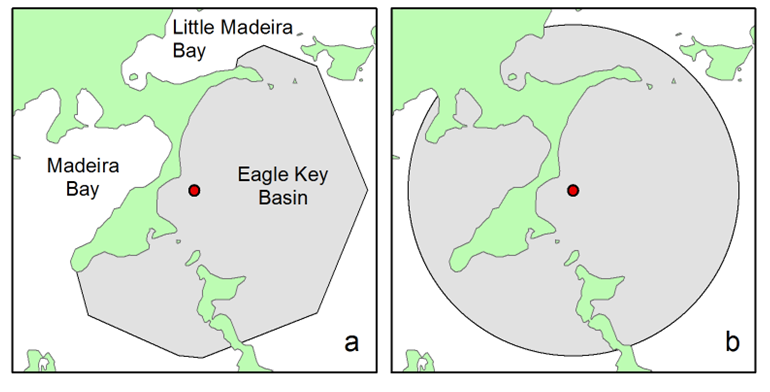}
\end{center}
\vspace{-20pt}
\caption{Interpolation neighborhood (shaded polygon) for a point in Eagle Key Basin (filled circle). Using inverse path distance weighting (a); Using inverse distance weighting (b). Note that in the case of inverse path distance weighting the interpolation neighborhood is limited by the cost-distance imposed by the land barrier.}
\label{fig:two}
\end{figure}
\FloatBarrier

\noindent A conventional application of path distance calculations might include route (road) planning between two points in a mountain range \citep{collischonn2000direction}. The analysis would seek to weight distances of various route alternatives by their accumulated travel "costs." These costs would be defined so that areas with extreme changes in elevation would be given a high travel cost.

  Ultimately, road construction might proceed along the route with the lowest cost-path. In the present study, land areas are given an impossibly high travel cost in order to restrict the interpolation neighborhood distances (d) to "in-water" distances. 
IPDW was carried out using the routines in the \pkg{ipdw} \proglang{R} package \citep{stachelek2014}. The order of operations to calculate spatial weights followed \citet{suominen2010surface}. Path distance calculations were computed following \citet{csardi2006igraph} and \citet{etten2014}. Reproducible examples demonstrating package functions can be found in the \pkg{ipdw} documentation (http://CRAN.r-project.org/package=ipdw). 

  The \pkg{ipdw} package computes path distances from each prediction point to each point in a set of measurement points. Path distances are calculated by tracking the accumulated cell-to-cell movement within an underlying cost raster. In this study, the cost raster was constructed by converting a vector shapefile representing the islands and shallow banks within Florida Bay and reclassifying open water and land areas to 1 and 10,000 respectively. The particular values assigned to open water and land areas were intended to ensure that the maximum cost path distance between interpolation and measurement points could not exceed the value assigned to land areas. Shapefiles were obtained from the Center for Spatial Analysis at the Florida Fish and Wildlife Conservation Commission (http://myfwc.com/research/gis/).  The cell size of the output raster was optimized in order to balance the need for the resolution of narrow barrier features against the limitations of computation time for the IPDW procedure. Optimization was performed by varying the size (grain) of the cost raster grid (50 – 100 m), calculating the edge density of water/land areas using the functions presented in \citet{vanderwal}, and visualizing this information using "scalograms" \citep{rutchey2009determining}. Scalograms are plots of cell size versus the value of a given landscape metric. A 60 meter cell size was chosen because there was a marked change in the slope of the scalograms after increasing the resolution from 60 to 70 m. 

  IPDW was performed on a subset of the full dataset because of the significant computation time associated with calculating path distances on the full dataset (> 6,000 measurements per Dataflow survey). Points were randomly selected with equal probability from each cell of a rectangular mesh (~35 points per cell, 1 point per 1.2 km2 grid cell) to form the training dataset. The remaining points form the validation dataset (discussed below). This spatially-balanced selection strategy has the advantage of creating a more regular (as opposed to clustered) data set. Interpolation accuracy has been shown to be greatly increased when regularly sampled datasets are available \citep{isaaks1989applied, zimmerman1999experimental}.  

  Interpolations were also performed using the standard Euclidean inverse distance weighting (IDW) routine in order to judge the benefit of using IPDW (Eq~\ref{eq:one}). IDW and IPDW have different interpolation neighborhoods by design. As a result, a given estimation point may be computed from a different overall number of neighbors. Given this fact, we kept all parameters (neighborhood size, power) the same during both procedures in order to standardize the way in which weights were assigned to measurements within the variously defined interpolation neighborhoods. Interpolation accuracy was determined using a cross validation procedure that compared interpolated predictions against the validation dataset. The output was used to calculate prediction mean absolute error (MAE) and root mean squared error (RMSE). Although MAE and RMSE provide similar information and are qualitatively similar, MAE is less sensitive to outliers. For this reason, we focus primarily on MAE as a diagnostic.

\section[Results]{Results}
\begin{figure}[h]
\begin{center}
\includegraphics{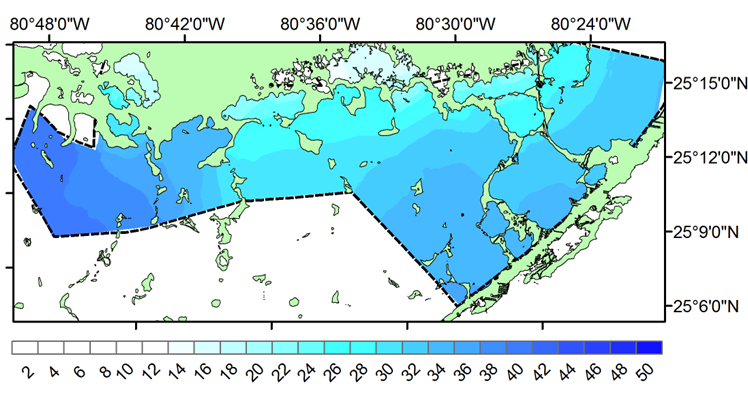}
\end{center}
\vspace{-20pt}
\caption{IPDW interpolated salinity averaged across all surveys from 2006-2012.}
\label{fig:three}
\end{figure}

\noindent A total of 23 Dataflow surveys were conducted from 2006 – 2012 (Table~\ref{table:one}). Salinities oscillated from an annual minimum at the end of the wet season (September-October) to an annual maximum in at the end of the dry season (May-June). Observed salinities varied from near fresh (<1) to hypersaline (>40). There was a general increasing trend in salinity with distance from the northeastern coastal embayments and there was a clear indication of a lower salinity "plume" extending through the middle of northeastern Florida Bay and Eagle Key basin (Figure~\ref{fig:two}, \ref{fig:three}). This is consistent with previous studies describing the general salinity patterns within Florida Bay \citep{kelble2007salinity}. A clear difference between past studies and IPDW interpolations is that the contribution of low salinity water from individual tributaries is clearly differentiable. For example, high salinities in Madeira Bay did not extend through the barrier separating it from Eagle Key Basin (Figure~\ref{fig:two}, \ref{fig:three}).

\vspace{20pt}
Table 1. Maximum, minimum, and range of measured salinities for each Dataflow survey
\begin{center}
\begin{tabular}{c c c c}
\hline
Date & Max & Min & Range \\
\hline
01-24-2006 & 33.99 & 11.73 & 22.26 \\
07-19-2006 & 40.18 & 10.83 & 29.35 \\
09-06-2006 & 28.60 & 0.19 & 28.41 \\
01-31-2007 & 35.03 & 18.57 & 16.46 \\
07-24-2007 & 36.40 & 1.25 & 35.15 \\
09-11-2007 & 42.86 & 3.54 & 39.32 \\
01-08-2008 & 34.62 & 14.39 & 20.23 \\
04-30-2008 & 51.22 & 17.14 & 34.08 \\
08-26-2008 & 40.40 & 6.83 & 33.57 \\
12-03-2008 & 34.68 & 7.54 & 27.14 \\
04-28-2009 & 52.34 & 32.86 & 19.48 \\
06-17-2009 & 48.09 & 9.21 & 38.88 \\ 
10-26-2009 & 40.27 & 1.40 & 38.87 \\
02-09-2010 & 33.25 & 1.18 & 32.07 \\
04-27-2010 & 41.04 & 9.15 & 31.89 \\
07-27-2010 & 41.97 & 2.40 & 39.57 \\
02-15-2011 & 30.37 & 1.16 & 29.21 \\
05-11-2011 & 48.83 & 14.83 & 34.00 \\
06-28-2011 & 48.67 & 25.39 & 23.28 \\
09-13-2011 & 38.03 & 2.72 & 35.31 \\
03-29-2012 & 38.60 & 16.13 & 22.47 \\
06-05-2012 & 37.19 & 0.00 & 37.19 \\
08-14-2012 & 39.33 & 0.44 & 38.89 \\
\hline
\label{table:one}
\end{tabular}
\end{center}
\vspace{20pt}

\begin{figure}[h]
\begin{center}
\includegraphics[width=100mm]{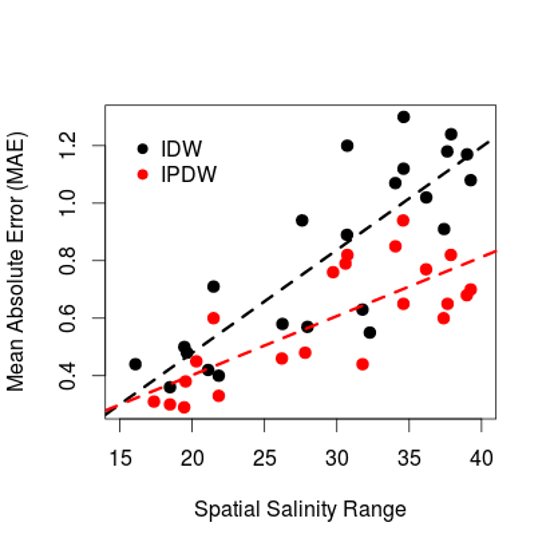}
\end{center}
\vspace{-20pt}
\caption{Comparison between prediction mean absolute error and the spatial salinity range for Dataflow surveys using IPDW (red points) and IDW (black points).}
\label{fig:five}
\end{figure}

\noindent In addition, to spatial variability, Florida Bay water quality is often subject to marked interannual variability (Table~\ref{table:one}). During an extremely dry period in April 2009, there was extensive hypersalinity (values greater than 40) extending from the bay into the northern embayments. In contrast, during an extremely wet period in April 2010, hypersalinity was more restricted. In general, the salinity transitions were more intense (abrupt) in April 2010 relative to April 2009. Interpolation error was higher in April 2010 coincident with these more abrupt salinity transitions. This illustrates a general trend whereby larger spatial salinity ranges were associated with higher interpolation error (Figure~\ref{fig:five}).
  
\FloatBarrier

\begin{figure}[h]
\begin{center}
\includegraphics{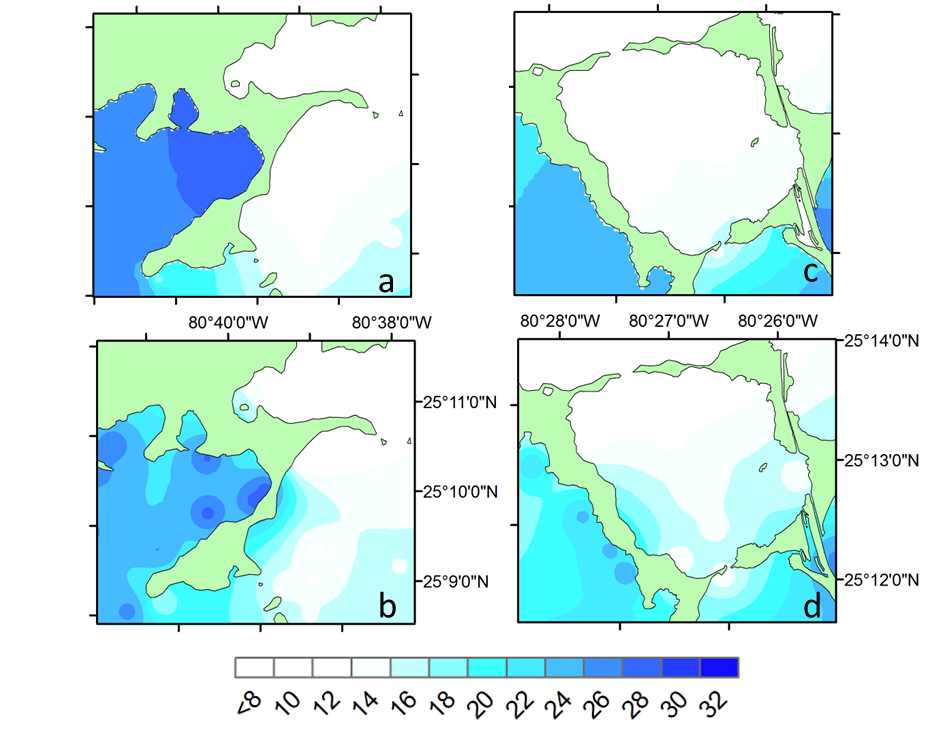}
\end{center}
\vspace{-20pt}
\caption{Locations from the August 2012 survey where IPDW produced notably different salinity results relative to IDW. Note that IPDW interpolations (a and c) account for barriers whereas IDW interpolations have unrealistically extended through barriers (b and d).}
\label{fig:four}
\end{figure}
\FloatBarrier

\noindent Generally, the IPDW procedure prevented underestimation of salinity on the downstream side of barriers and overestimation of salinity on the upstream side (Figure~\ref{fig:four}). The greatest improvements in accuracy were found in the western portions of Florida Bay where hypersaline basins are narrowly separated from brackish water embayments by multiple complex land barriers.   

  On a system-wide basis, the IPDW procedure predicted salinity values with a range of RMSE between 0.5-1.87 and a MAE between 0.29-0.94. Comparative IDW interpolations predicted salinity values with a range of RMSE between 0.6-2.19 and a MAE between 0.36-1.3.  System-wide prediction error (MAE and RMSE) was significantly higher using the IDW procedure relative to IPDW (Wilcoxon signed rank test, p < 0.01).  Lower IPDW prediction errors were more evident in specific basins such as Little Blackwater Sound (Figure~\ref{fig:four}b, \ref{fig:four}d). The range of MAE for Little Blackwater Sound was 0.14-1.92 and 0.13-5.43 for IPDW and IDW respectively. IDW prediction errors for Little Blackwater Sound were significantly higher than IPDW prediction errors (Wilcoxon signed rank test, p < 0.01).

\FloatBarrier

\section[Discussion]{Discussion}
\noindent In this study, our non-Euclidean interpolation (IPDW) method provided increased accuracy and resolution of water quality features relative to Euclidean interpolation (IDW). This increased accuracy was apparent in two ways. First, the results of both procedures showed that higher interpolation errors (MAE) were associated with surveys when there was a larger range of measured salinities across the study area. As the range of salinities increased, spatial gradients became more intense, and the benefits of IPDW became more evident (Figure~\ref{fig:five}). The ability to recover the location and shape of intense spatial gradients is a unique benefit of combining high density sampling with non-Euclidean interpolation.  The second way in which IPDW was beneficial was that it preserved the separation between nearshore embayments where water bodies with distinct water quality characteristics are located in close proximity to one another (Figure~\ref{fig:four}). These embayments are of great interest because they are sensitive to variable freshwater discharge from the Everglades \citep{nuttle2000influence}. Close tracking of their condition is necessary in order to evaluate the impact of environmental restoration projects designed to increase freshwater discharges to Florida Bay. Many of these projects are part of the Comprehensive Everglades Restoration Plan (CERP).

  However, in the open areas of Florida Bay, there was little improvement in accuracy. This finding is consistent with \cite{suominen2010surface} where interpolations were performed across an area with few contiguous barriers and non-Euclidean methods provided little benefit. In contrast, \cite{greenberg2011least} showed marked improvements in accuracy using non-Euclidean interpolation within a stream network characterized by contiguous barriers. Overall, non-Euclidean methods are likely to provide the most benefit when the spatial domain is bisected by contiguous barriers. In the absence of these barriers, Euclidean methods may be sufficient \citep{suominen2010surface,rivera2011salinity}. 

  In addition to the presence of contiguous barriers within the spatial domain, there are several important considerations to be weighed prior to implementing IPDW. First, non-Euclidean "path" distances could be used as input to a variety of geostatistical methods. Throughout this study, we use our ipdw R package which takes path distances as input to inverse distance weighting. Alternative applications that use path distances as input to kriging are likely to be more powerful but have some potential pitfalls (see discussion in \citep{lopez2009geostatistical}. \citet{lopez2009geostatistical} found path distance based kriging to be effective while at the same time providing estimation of prediction errors. As software becomes available, future studies will examine path distance based interpolation of spatial data using a geostatistical approach (kriging). A second consideration is that interpolation with path distances can be computationally demanding and difficult to implement. This can be attributed to resource intensive nature of path distance calculations and the lack of built-in routines in GIS software available outside of a scripting environment \citep{stachelek2014}. 

  This is the first study to combine non-Euclidean interpolation with high density Dataflow sampling. Our methodology could be used to expand the spatial domain of interpolations which have been limited in scope to along the survey track itself \citep{morse2011environmental, soderqvist2010seasonal, xie2013geographically}. It should be noted, however, that the process of subsetting the full dataset based on a grid will only provide a more regularly spaced dataset when the survey track has followed a sufficient number of meanders. Subsetting linear survey tracks such as those in \cite{buzzelli2014fine} may be necessary in order to satisfy the limitations to computation but will not provide a more regularly sampled dataset. An alternative to our methodology is to remap the coordinates of measurement points into a new Euclidean space that accounts for "in-water" distances \citep{loland2003spatial}. This may allow for the use of traditional Euclidean interpolation methods while still accounting for barriers. Generating this transformed output adds an additional level of complexity to processing routines. 

  Ultimately the ability of interpolation procedures to recover spatial water quality patterns is a function of the underlying monitoring program design. To our knowledge, non-Euclidean distance calculations have yet to be applied to the design of coastal water quality sampling. Potential applications include combining Dataflow and IPDW to examine optimal spacing of measurement points. For example, \cite{anttila2008feasible} used dense geospatial output to establish that the spatial representativeness (range) of single point chlorophyll measurements was quite low. This may be particularly informative given that studies examining non-Euclidean water quality interpolation have typically used sparse, spatially distributed sampling designs rather than spatially dense data collection \citep{greenberg2011least, suominen2010surface}. 

  Interpolated Dataflow output has a variety of potential applications. On a basic level, accurate interpolations provide a spatially intensive snapshot of the water quality conditions. This is especially useful in evaluating the state of the system and documenting the response to extreme events such as hurricanes and droughts \citep{davis2004importance}. More applied uses include the validation of remote sensing \citep{xie2013geographically}, ecological \citep{fourqurean2003forecasting, madden2009florida}, statistical \citep{marshall2011empirical}, and hydrologic models \citep{nuttle2000influence}. For future studies, researchers should consider using IPDW or related non-Euclidean interpolation methods, given the proliferation of software tools, potential improvements in accuracy, and the more widespread adoption of Dataflow technology.

\bibliography{stachmad}
\end{document}